\documentclass[11pt]{article}
\usepackage{amssymb}
\usepackage{graphics}
\usepackage{epsfig}
\usepackage{a4wide}
\begin{document}

\title{ Pair Production of Charged Higgs Bosons from Bottom-Quark Fusion } \vspace{3mm}

\author{{ Hou Hong-Sheng$^{2}$, Ma Wen-Gan$^{1,2}$, Zhang Ren-You$^{2}$, Jiang Yi$^{2}$, Han Liang$^{2}$, Xing Li-Rong$^{2}$}\\
{\small $^{1}$CCAST (World Laboratory), P.O.Box 8730, Beijing, 100080, People's Republic of China} \\
{\small $^{2}$Department of Modern Physics, University of Science and Technology of China (USTC),}\\
{\small       Hefei, Anhui 230026, People's Republic of China} }

\date{}
\maketitle
\vskip 12mm

\begin{abstract}
 For very large values of $\tan\beta$, the charged Higgs
boson pair production via $b\bar{b}$ annihilation can proceed
dominantly at the Large Hadron Collider (LHC) . We calculated the
cross sections of the charged Higgs boson pair production via
subprocess $b\bar{b} \to H^+H^-$ at the LHC including the
next-to-leading order (NLO) QCD corrections in the minimal
supersymmetric standard model (MSSM). We find that the NLO QCD
corrections can significantly reduce the dependence of the cross
sections on the renormalization and factorization scales.
\par
\end{abstract}

\vskip 5cm


\vfill \eject

\baselineskip=0.32in

\renewcommand{\theequation}{\arabic{section}.\arabic{equation}}
\renewcommand{\thesection}{\Roman{section}.}
\newcommand{\nb}{\nonumber}

\makeatletter      
\@addtoreset{equation}{section}
\makeatother       

\par
\section{Introduction}
One of the most important missions of future high-energy
experiments is to search for scalar Higgs bosons and explore the
electroweak symmetry breaking mechanism. In the standard model
(SM)\cite{sm}, one doublet of complex scalar fields is needed to
spontaneously break the symmetry, leading to a single neutral
Higgs boson $h^0$. The minimal supersymmetric standard model
(MSSM) \cite{mssm} is one of the most attractive extensions of the
SM. The MSSM requires the existence of two doublets of Higgs
fields to cancel anomalies and to give masses separately to up and
down-type fermions. The MSSM predicts two CP-even neutral Higgs
bosons $h^0,H^0$, a pseudoscalar $A^0$ Higgs boson and a pair of
charged scalar particles $H^{\pm}$. At the tree level, the MSSM
Higgs sector has two free parameters: $\tan\beta = v_2/v_1$, the
ratio of the vacuum expectation values of the two Higgs doublets
and a Higgs boson mass which is taken to be $m_{H^{\pm}}$ in this
paper.
\par
The discovery of the $H^{\pm}$ would be a clear signal for the
existence of physics beyond the SM with a strong hint towards
supersymmetry. The CERN large hadron collider (LHC), with
$\sqrt{s}=14~TeV$ and a luminosity of 100 ${\rm fb}^{-1}$ per
year, will be a wonderful tool for looking for new physics. At the
LHC, the light charged Higgs boson can be produced from the top
quark decays $t\to b + H^+$\cite{htb}. Heavy charged Higgs boson
is mainly produced via the processes $g\bar{b} \to
\bar{t}H^+$\cite{gbht}, $gg \to \bar{t} b H^+$\cite{gghtb} and $q
b \to q'bH^+$\cite{qqhtb}. Moreover, single charged Higgs boson
production associated with a $W$ boson, via tree-level $b\bar{b}$
annihilation and one-loop $gg$ fusion, has been proposed and
analyzed in Ref.\cite{bbhw}.
\par
At the LHC, the charged Higgs boson also can be produced in pair
production mode. There are three important $H^+H^-$ production
channels: (i) $q\bar{q} \to H^+H^-$, where $q = u, d, c, s, b$.
(via Drell-Yan process, where a photon and a Z-boson are exchanged
in the $s$-channel. In the case of $q = b$, there are additional
Feynman diagrams involving $h^0$ and $H^0$ in the $s$-channel and
the top quark in the $t$-channel) \cite{qqhh}. For very large
values of $\tan\beta$, due to the large contributions from the
additional diagrams, the $H^+H^-$ production can proceed
dominantly via $b\bar{b}$ annihilation \cite{gghh1}. (ii) $gg \to
H^+H^-$ (via quarks and squarks loop) \cite{gghh1}\cite{gghh} .
(iii) $qq \to qqH^+H^-$ (via vector boson fusion) \cite{vvhh}.
\par
In the subprocess $b\bar{b} \to H^+H^-$, the initial state bottom
quarks arise from a gluon in the proton splitting into a collinear
$b\bar{b}$-pair, parameterized in terms of bottom quark
distribution functions. On the other hand, the 'twin' process $gg
\to b\bar{b}H^+H^-$ using gluon density has been studied at LO in
Ref.\cite{gghhbb}. It is pointed out that the use of the b-quark
density may overestimate the inclusive cross section due to crude
approximations in the kinematics \cite{wrongb}. However, it is
suggested that the bottom quark parton approximation maybe valid
by choosing appropriate factorization scale \cite{factorscale}.
Following the suggestions in Ref.\cite{wrongb,factorscale}, we can
analyze the transverse momentum distribution of the b-quarks in
the process $gg \to b \bar{b} H^+ H^-$ as shown in Fig.2 of
Ref.\cite{gghhbb}. The most suitable factorization scale for
$b\bar{b} \to H^+H^-$ is of order $m_{H^\pm}/5 \sim m_{H^\pm}/4$
which is much smaller than the usually used scale $m_{H^\pm}$.
\par
In this paper, we study the process $pp \to b\bar{b} \to H^+H^-+X$
at the LHC with very large $\tan\beta$ in the MSSM. The NLO QCD
corrections are calculated. The paper is organized as follow: In
section 2, we discuss the LO results of the subprocess $b\bar{b}
\to H^+H^-$. In section 3, we present the calculations of the NLO
QCD corrections. In section 4, the numerical results, discussions
and conclusions are presented.

\vskip 5mm
\section{The Leading Order Cross Section}
The Feynman diagrams for the subprocess $b(p_1)\bar{b}(p_2) \to
H^+(k_3)H^-(k_4)$ at the LO are shown in Fig.1, where $p_{1,2}$
and $k_{3,4}$ represent the four-momenta of the incoming partons
and the outgoing particles respectively.

\begin{figure}[htbp]
\centering
\scalebox{0.75}{\includegraphics*[100,547][640,650]{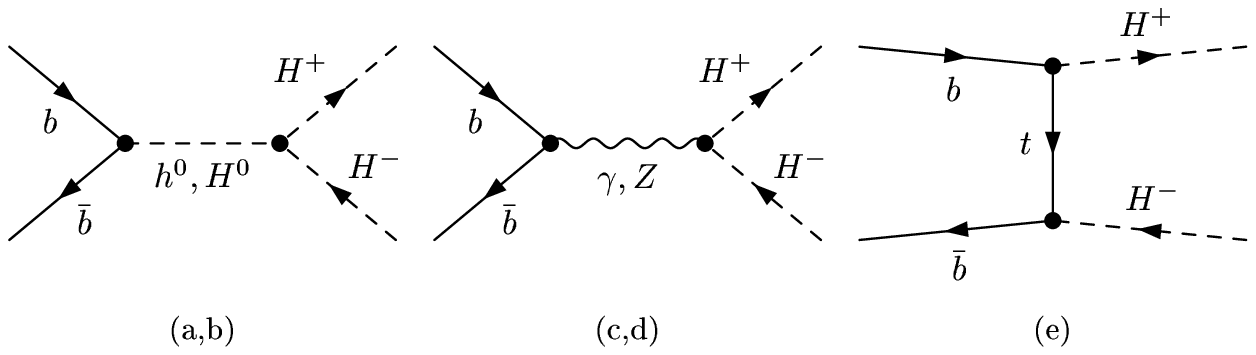}}
\caption{The tree level Feynman diagrams for $b \bar{b} \to H^+
H^-$.}
\end{figure}

We divide the tree-level amplitude into two parts,
\begin{eqnarray}
M^0=M_0^{(s)}+M_0^{(t)},
\end{eqnarray}
where $M_0^{(s)}$ and $M_0^{(t)}$ represent the amplitudes arising
from the s-channel diagrams shown in Fig.1(a,b,c,d) and the
t-channel diagram shown in Fig.1(e) respectively. The explicit
expressions for the amplitudes $M_0^{(s)}$ and $M_0^{(t)}$ can be
written as
\begin{eqnarray}
M_0^{(t)}&=& i\bar{v}(p_2) \frac{(P_R g^{(1)}_{H^+tb}+P_L
g^{(2)}_{H^+tb}) (m_t+ \rlap/p_1 - \rlap/k_3)(P_L
g^{(1)}_{H^+tb}+P_R g^{(2)}_{H^+tb})}{\hat{t}-m^2_t}
u(p_1), \nb \\
M_0^{(s)}&=& i \bar{v}(p_2) [
\frac{g_{H^+H^-h^0}g_{h^0b\bar{b}}}{\hat{s}-m^2_{h^0}}
+\frac{g_{H^+H^-H^0}g_{H^0b\bar{b}}}{\hat{s}-m^2_{H^0}} -\frac{Q_b
e^2(\rlap/k_3 - \rlap/k_4) }{\hat{s}} \nb \\
&-&\frac{(c^2_w-s^2_w)e^2}{2 c^2_w s^2_w} \frac{(\rlap/k_3 -
\rlap/k_4) (s^2_w/3-P_L/2)}{\hat{s}-m^2_Z} ] u(p_1),
\end{eqnarray}
where $\hat{s}=(p_1+p_2)^2$, $\hat{t}=(p_1-k_3)^2$ and
$\hat{u}=(p_1-k_4)^2$ are the usual Mandelstam variables.
$P_{L,R}=(1 \mp \gamma_5)/2$, $s_w=\sin \theta_w$, $c_w=\cos
\theta_w$,$Q_b = -1/3$. The couplings are defined below,
\begin{eqnarray}
g^{(1)}_{H^+tb} &=& \frac{i e m_t}{\sqrt{2} m_W s_w \tan \beta},
~~~~~~g^{(2)}_{H^+tb} = \frac{i e \overline{m}_b (\mu_r) \tan
\beta}{\sqrt{2} m_W s_w }, \nb \\
g_{h^0b\bar{b}} &=& \frac{i e \overline{m}_b (\mu_r)}{2 m_W s_w}
\frac{\sin \alpha}{\cos \beta},~~~~~~g_{H^0b\bar{b}} = - \frac{i e
\overline{m}_b (\mu_r)}{2 m_W s_w} \frac{\cos \alpha}{\cos \beta},
\nb \\
g_{H^+H^-h^0} &=& -\frac{i e m_W}{s_w} (\sin(\beta-\alpha)+
\frac{\cos(2 \beta) \sin(\alpha+\beta)}{2 c^2_w}), \nb \\
g_{H^+H^-H^0} &=& -\frac{i e m_W}{s_w} (\cos(\beta-\alpha)-
\frac{\cos(2 \beta) \cos(\alpha+\beta)}{2 c^2_w}),
\end{eqnarray}
where $\alpha$ is the mixing angle which leads to the physical
Higgs eigenstates $h^0$ and $H^0$. $\overline{m}_b(\mu_r)$ is
$\overline{\rm MS}$ running mass of the bottom quark. We neglected
the bottom quark mass during our calculation except in the Yukawa
couplings.
\par
Then the LO cross section for the subprocess $b\bar{b} \to H^+H^-$
is obtained by using the following formula:
\begin{eqnarray}
\label{folding} \hat{\sigma}^0(\hat{s}, b\bar{b} \to H^+H^-) =
\frac{1}{16 \pi \hat{s}^2} \int_{\hat{t}_{min}}^{\hat{t}_{max}}
d\hat{t}~ \overline{\sum} |M^0|^2,
\end{eqnarray}
where $\hat{t}_{max,min}= (m_{H^\pm}- \frac{1}{2} \hat{s}) \pm
\frac{1}{2}\sqrt{\hat{s}^2- 4 m^2_{H^\pm} \hat{s}} $. The
summation is taken over the spins and colors of initial and final
states, and the bar over the summation denotes averaging over the
spins and colors of initial partons.

\vskip 5mm
\section{NLO QCD Corrections}

The NLO QCD corrections to $pp \to b\bar{b} \to H^+H^-+X$ in the
MSSM can be separated into two parts: the virtual corrections
arising from one loop diagrams and the real corrections.
\par
\subsection{Virtual One-loop Corrections }

The virtual one-loop diagrams of the subprocess $b\bar{b} \to
H^+H^- $ in the MSSM consist of self-energy, vertex and box
diagrams which are depicted in Figs.2-3. Fig.2 shows the diagrams
of the SM-like QCD corrections arising from quark and gluon loops,
and Fig.3 shows the diagrams of the so called 'pure' SUSY QCD
corrections arising from squark and gluino loops. There exist both
ultraviolet(UV) and soft/collinear infrared(IR) singularities in
the amplitude for the SM-like diagrams shown in Fig.2. The
amplitude for the 'pure' SUSY QCD diagrams (Fig.3) contains only
UV singularities. In our calculation, we adopt the 't
Hooft-Feynman gauge and all the divergences are regularized by
using dimensional regularization method in $d=4-2 \epsilon$
dimensions.

\begin{figure}[htbp]
\centering
\scalebox{0.75}{\includegraphics*[100,440][640,640]{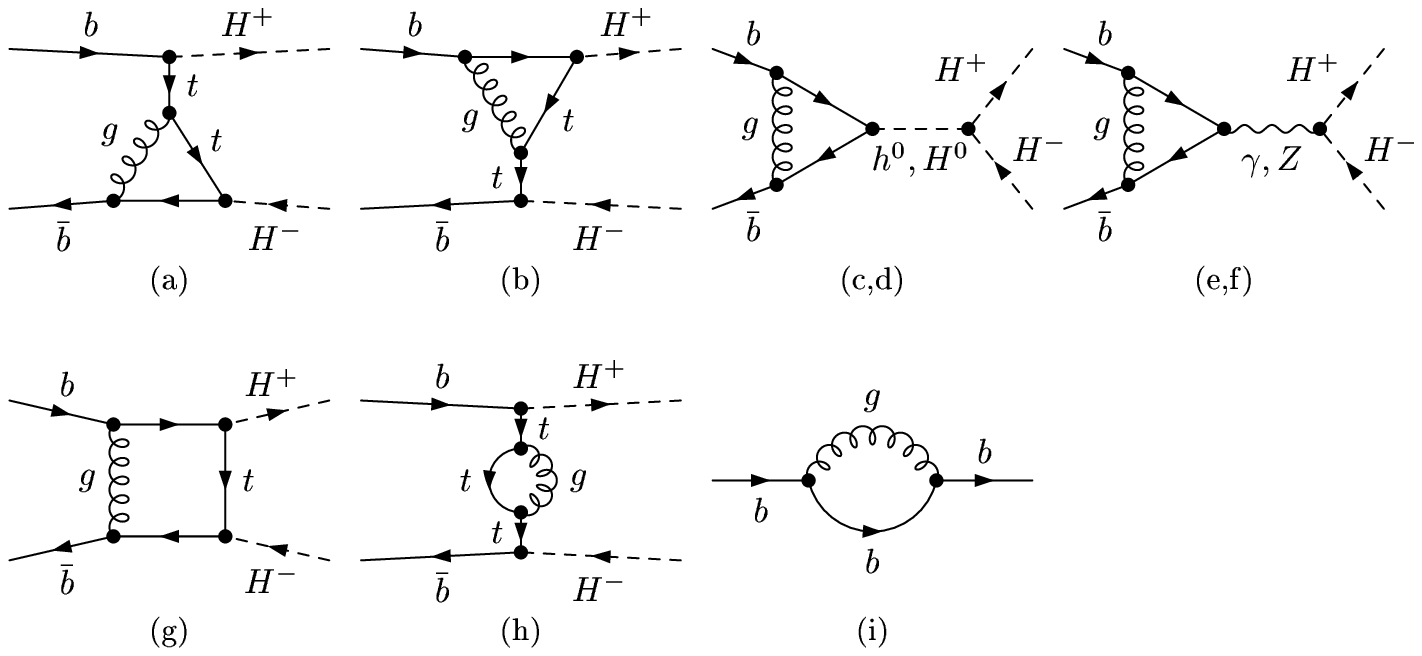}}
\caption{The one-loop Feynman diagrams of the SM-like QCD
corrections for $b \bar{b} \to H^+ H^-$ subprocess.}
\end{figure}

\begin{figure}[htbp]
\centering
\scalebox{0.75}{\includegraphics*[100,440][640,630]{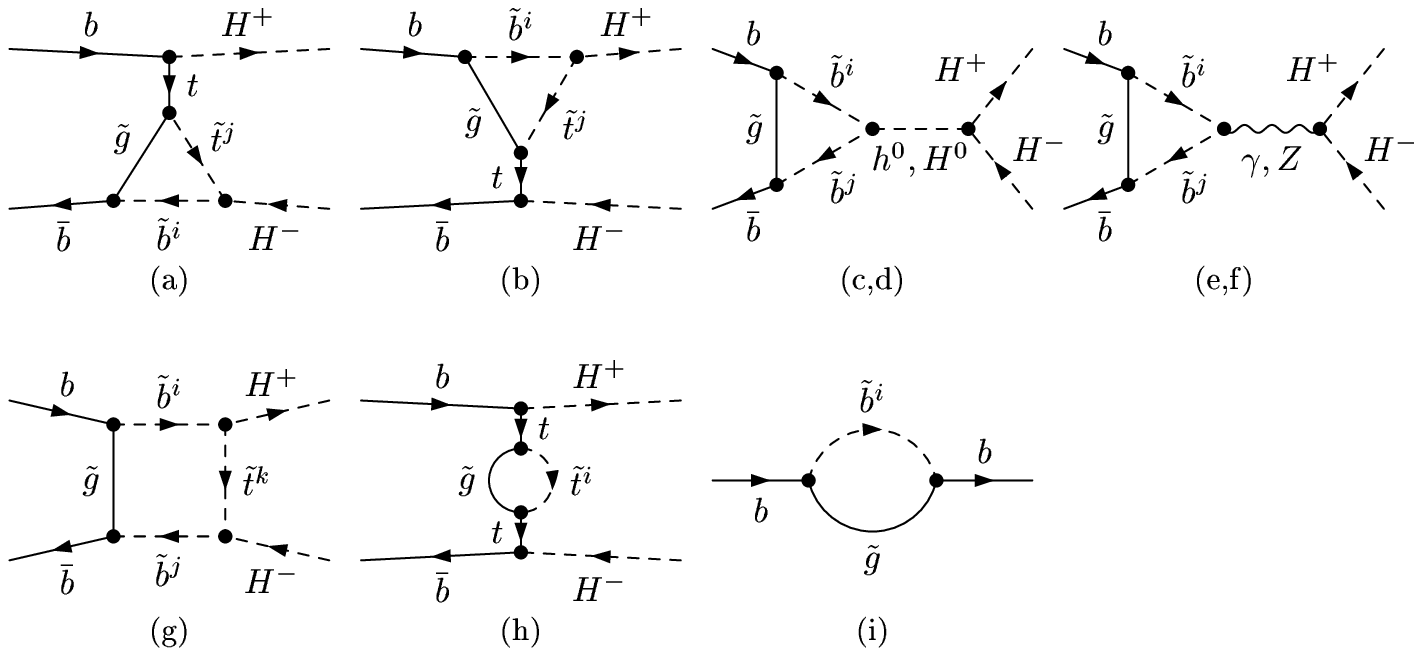}}
\caption{The one-loop Feynman diagrams of the 'pure' SUSY QCD
corrections for $b \bar{b} \to H^+ H^-$ subprocess.}
\end{figure}

\par
In order to remove the UV divergences, we need to renormalize the
wave functions of the external fields and the Yukawa couplings of
$h^0-b-\bar{b}$, $H^0-b-\bar{b}$ and $H^--t-\bar{b}$. We
renormalize the top quark mass in the on-mass-shell (OS) scheme.
For the renormalization of the bottom quark mass in the Yukawa
couplings, we employ the modified minimal subtraction
($\overline{\rm MS}$) scheme. The relevant renormalization
constants in this work can be expressed as
\begin{eqnarray}
\label{renormalization constants} \frac{\delta m_t}{m_t} &=&
(\frac{\delta m_t}{m_t})^{(QCD)}+(\frac{\delta
m_t}{m_t})^{(SQCD)}\nb \\ &=& -\frac{\alpha_s}{4 \pi} C_F\{(4 B_0
+2 B_1)(m^2_t,m^2_t,0)-1\} \nb \\ &-&\frac{\alpha_s}{4 \pi}
C_F\{\sum^{2}_{i=1} [B_1 -
\frac{m_{\tilde{g}}}{m_t}\sin(2\theta_{\tilde{t}})(-1)^i
B_0](m^2_t,m^2_{\tilde{g}},m^2_{\tilde{t}_i}) \} ,\nb \\
\delta Z^L_b &=& (\delta Z^L_b)^{(QCD)}+(\delta Z^L_b)^{(SQCD)} \nb \\
&=& - \frac{\alpha_s}{4 \pi} C_F B_0(0,0,0)+ \frac{\alpha_s}{2
\pi} C_F \left[
B_1(0,m^2_{\tilde{g}},m^2_{\tilde{b}_1})\cos^2\theta_{\tilde{b}}+
B_1(0,m^2_{\tilde{g}},m^2_{\tilde{b}_2})\sin^2\theta_{\tilde{b}}\right], \nb \\
\delta Z^R_b &=& (\delta Z^R_b)^{(QCD)}+(\delta Z^R_b)^{(SQCD)} \nb \\
&=& - \frac{\alpha_s}{4 \pi} C_F B_0(0,0,0)+ \frac{\alpha_s}{2
\pi} C_F \left[
B_1(0,m^2_{\tilde{g}},m^2_{\tilde{b}_1})\sin^2\theta_{\tilde{b}}+
B_1(0,m^2_{\tilde{g}},m^2_{\tilde{b}_2})\cos^2\theta_{\tilde{b}}\right], \nb \\
\frac{\delta m_b}{m_b} &=& (\frac{\delta
m_b}{m_b})^{(QCD)}+(\frac{\delta m_b}{m_b})^{(SQCD)}\nb \\
&=& -\frac{\alpha_s}{4 \pi} 3 C_F \Delta +\frac{\alpha_s}{4 \pi}
C_F \Delta,
\end{eqnarray}
where $\Delta=\frac{1}{\epsilon} -\gamma_E + \ln(4 \pi)$ and
$C_F=\frac{4}{3}$. In above equations we divide the
renormalization constants into two parts, one arises from the
one-loop diagrams involving quark and gluon, the other comes from
the loops involving squark and gluino.

\par
The virtual corrections to the cross section for the subprocess
$b\bar b \to H^+H^-$ can be written as
\begin{eqnarray}
\hat{\sigma}^{V}(\hat{s}, b\bar{b} \to H^+H^-) = \frac{1}{16 \pi
\hat{s}^2} \int_{\hat{t}_{min}}^{\hat{t}_{max}} d\hat{t}~ 2 Re
\overline{\sum} [(M^{V})^{\dagger} M^0],
\end{eqnarray}
where $\hat{t}_{max,min}= (m_{H^\pm}- \frac{1}{2} \hat{s}) \pm
\frac{1}{2}\sqrt{\hat{s}^2- 4 m^2_{H^\pm} \hat{s}} $, and the
summation with bar over head means the same operation as that
appeared in Eq.(\ref{folding}). $M^{V}$ is the renormalized
amplitude for virtual one-loop corrections. After renormalization
procedure, $\hat{\sigma}^{V}$ is UV-finite. Nevertheless, it still
contains the soft/collinear IR singularities
\begin{eqnarray}
\label{virtual cross section}
d\hat{\sigma}^V|_{IR}=\left[\frac{\alpha_s}{2 \pi}
\frac{\Gamma(1-\epsilon)}{\Gamma(1-2 \epsilon)}\left(\frac{4 \pi
\mu_r^2}{\hat{s}}\right)^{\epsilon}\right]d\hat{\sigma}^0
\left(\frac{A^V_2}{\epsilon^2}+\frac{A^V_1}{\epsilon} \right),
\end{eqnarray}
where
\begin{eqnarray}
A^V_2&=& -2 C_F, ~~~~~~~A^V_1 = -3 C_F.
\end{eqnarray}

\par
The soft divergences will be cancelled by adding the real gluon
emission corrections. The remaining collinear divergences can be
absorbed into the parton distribution functions, which will be
discussed in the following subsections.

\subsection{Real Gluon Emission Corrections of $b\bar{b} \to H^+H^-+g$ }

The real gluon emission subprocess $b\bar{b} \to H^+H^-+g$ (shown
in Fig.4) presents ${\cal O}(\alpha_s)$ correction to $b\bar{b}
\to H^+H^-$. It also gives the IR singularities which cancel the
analogous singularities arising from the one-loop level virtual
corrections mentioned in the above subsection. These singularities
can be either of soft or collinear nature and can be conveniently
isolated by slicing the phase space of subprocess $b\bar{b} \to
H^+H^-+g$ into different regions defined by suitable cutoffs, a
method which goes under the general name of the phase space
slicing method (PPS)\cite{PSS1}.

\begin{figure}[htbp]
\centering
\scalebox{0.75}{\includegraphics*[100,440][640,650]{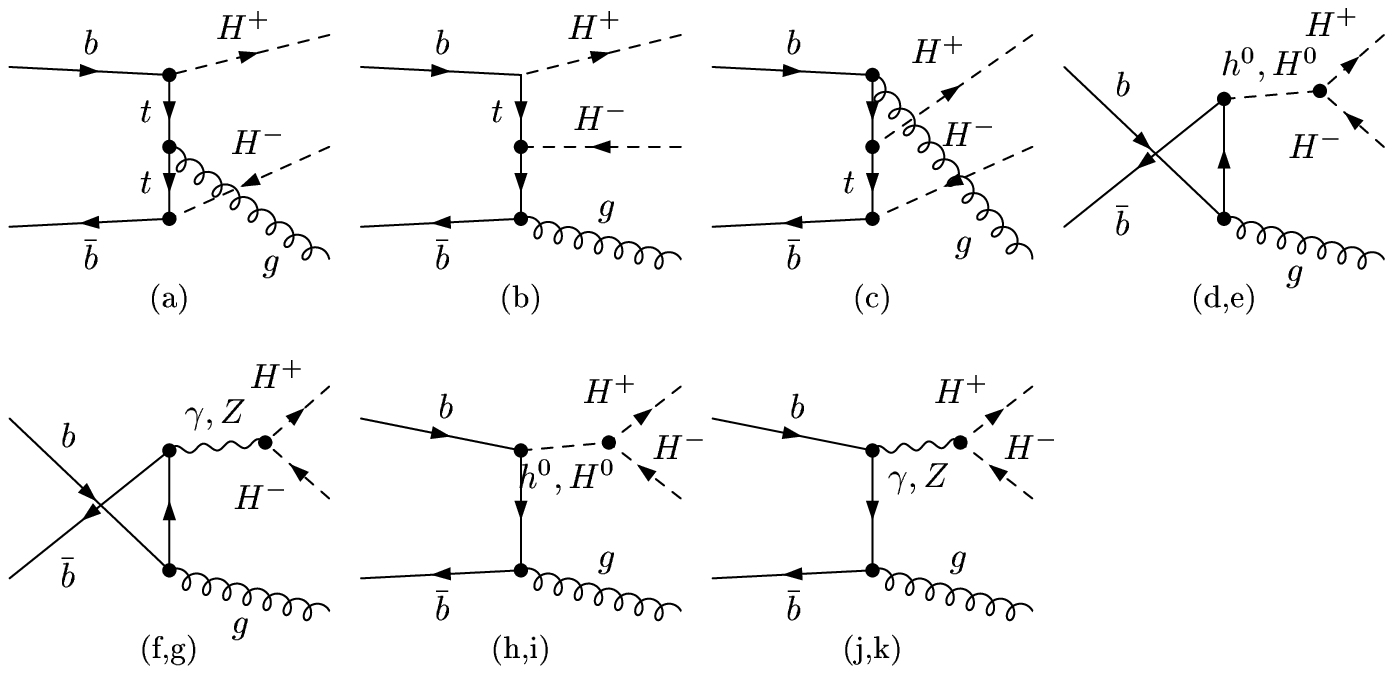}}
\caption{The Feynman diagrams for $b \bar{b} \to H^+ H^- +g$.}
\end{figure}
\par
We denote this $2 \to 3$ subprocess as
\begin{eqnarray}
b(p_1)+ \bar{b}(p_2) \to H^+(k_3)+H^-(k_4)+g(k_5)
\end{eqnarray}
and calculate the cross section by using the two cutoff phase
space slicing method\cite{Harris}. We define the Lorentz
invariants
\begin{eqnarray}
\hat{s}&=&(p_1+p_2)^2,~~\hat{t}=(p_1-k_3)^2,~~\hat{u}=(p_1-k_4)^2,\nb
\\ \hat{t}_{15}&=&(p_1-k_5)^2,~~ \hat{t}_{25}=(p_2-k_5)^2
\end{eqnarray}
and describe this method briefly as follows: Firstly, by
introducing an arbitrary small soft cutoff $\delta_s$ we separate
the $2 \to 3$ phase space into two regions, according to whether
the energy of the emitted gluon is soft, i.e. $E_5 \leq
\delta_s\sqrt{\hat{s}}/2$, or hard, i.e. $E_5 >
\delta_s\sqrt{\hat{s}}/2$. The partonic real cross section can be
written as
\begin{eqnarray}
\hat{\sigma}^R_g(b\bar{b} \to H^+H^-g)=\hat{\sigma}^S_g(b\bar{b}
\to H^+H^-g)+\hat{\sigma}^H_g(b\bar{b} \to H^+H^-g),
\end{eqnarray}
where $\hat{\sigma}^S_g$ is obtained by integrating over the soft
region of the emitted gluon phase space. $\hat{\sigma}^S_g$
contains all the soft IR singularities. Secondly, to isolate the
remaining collinear singularities from $\hat{\sigma}^H_g$, we
further decompose $\hat{\sigma}^H_g$ into a sum of hard collinear
(HC) and hard non-collinear ($\overline{\rm HC}$) terms by
introducing another cutoff $\delta_c$ named collinear cutoff
\begin{eqnarray}
\hat{\sigma}^H_g(b\bar{b} \to H^+H^-g)=\hat{\sigma}^{\rm
HC}_g(b\bar{b} \to H^+H^-g)+\hat{\sigma}^{\overline{\rm
HC}}_g(b\bar{b} \to H^+H^-g).
\end{eqnarray}
The HC regions of the phase space are those where any one of the
Lorentz invariants $\hat{t}_{15},\hat{t}_{25}$ becomes smaller in
magnitude than $\delta_c \hat{s} $, while at the same time the
emitted gluon remains hard. $\hat{\sigma}^{\rm HC}_g$ contains the
collinear divergences. In the soft and HC region,
$\hat{\sigma}^S_g$ and $\hat{\sigma}^{\rm HC}_g$ can be obtained
by performing the phase space integration in $d$-dimension
analytically. In the $\overline{\rm HC}$ region,
$\hat{\sigma}^{\overline{\rm HC}}_g$ is finite and can be
evaluated in four dimensions using standard Monte Carlo
techniques\cite{Lepage}. The cross sections, $\hat{\sigma}^S_g$,
$\hat{\sigma}^{\rm HC}_g$ and $\hat{\sigma}^{\overline{\rm
HC}}_g$, depend on the two arbitrary parameters, $\delta_s$ and
$\delta_c$. However, in the total real gluon emission hadronic
cross section $\sigma^R_g$, after mass factorization, the
dependence on these arbitrary cutoffs cancels, as will be
explicitly shown in Sec. 4. This constitutes an important check of
our calculation.
\par
The differential cross section in the soft region is given as
\begin{eqnarray}
\label{soft cross section} d\hat{\sigma}^S_g=d\hat{\sigma}^0
\left[\frac{\alpha_s}{2 \pi} \frac{\Gamma(1-\epsilon)}{\Gamma(1-2
\epsilon)}\left(\frac{4 \pi
\mu_r^2}{\hat{s}}\right)^{\epsilon}\right]
\left(\frac{A^S_2}{\epsilon^2}+\frac{A^S_1}{\epsilon}+A^S_0
\right),
\end{eqnarray}
with
\begin{eqnarray}
A^S_2&=& 2 C_F,~~~A^S_1 = - 4 C_F \ln \delta_s, ~~~A^S_0 = 4 C_F
\ln^2 \delta_s.
\end{eqnarray}
\par
The differential cross section $d\sigma^{\rm HC}_g$ can be written
as
\begin{eqnarray}
\label{collinear-g} d\sigma^{\rm HC}_g&=&d\hat{\sigma}^0
\left[\frac{\alpha_s}{2 \pi} \frac{\Gamma(1-\epsilon)}{\Gamma(1-2
\epsilon)}\left(\frac{4 \pi
\mu_r^2}{\hat{s}}\right)^{\epsilon}\right]
(-\frac{1}{\epsilon})\delta_c^{-\epsilon} [2
P_{bb}(z,\epsilon)G_{b/P}(x_1/z)G_{\bar{b}/P}(x_2) \nb \\
&+& (x_1\leftrightarrow
x_2)]\frac{dz}{z}(\frac{1-z}{z})^{-\epsilon}dx_1dx_2.
\end{eqnarray}
where $G_{b,\bar{b}/P}(x)$ is the bare parton distribution
function of $b(\bar b)$ quark in proton. $P_{bb}(z,\epsilon)$ is
the $d$-dimensional unregulated ($z<1$) splitting function related
to the usual Altarelli-Parisi splitting kernel \cite{Altarelli}.
$P_{bb}(z,\epsilon)$ can be written explicitly as
\begin{eqnarray}
P_{bb}(z,\epsilon)&=&P_{bb}(z)+ \epsilon P'_{bb}(z), \nb \\
P_{bb}(z)&=&C_F \frac{1+z^2}{1-z},~~~~~~~~ P'_{bb}(z)=-C_F (1-z).
\end{eqnarray}

\subsection{Real Corrections from subprocesses $g b(\bar{b}) \to H^+H^-+b(\bar{b})$ }

In addition to the real gluon emission subprocess $b\bar{b} \to
H^+H^-+g$, there are also subprocesses $g b(\bar{b}) \to
H^+H^-+b(\bar{b})$ at this order of perturbation theory, as shown
in Fig.5.

\begin{figure}[htbp]
\centering
\scalebox{0.75}{\includegraphics*[100,440][640,650]{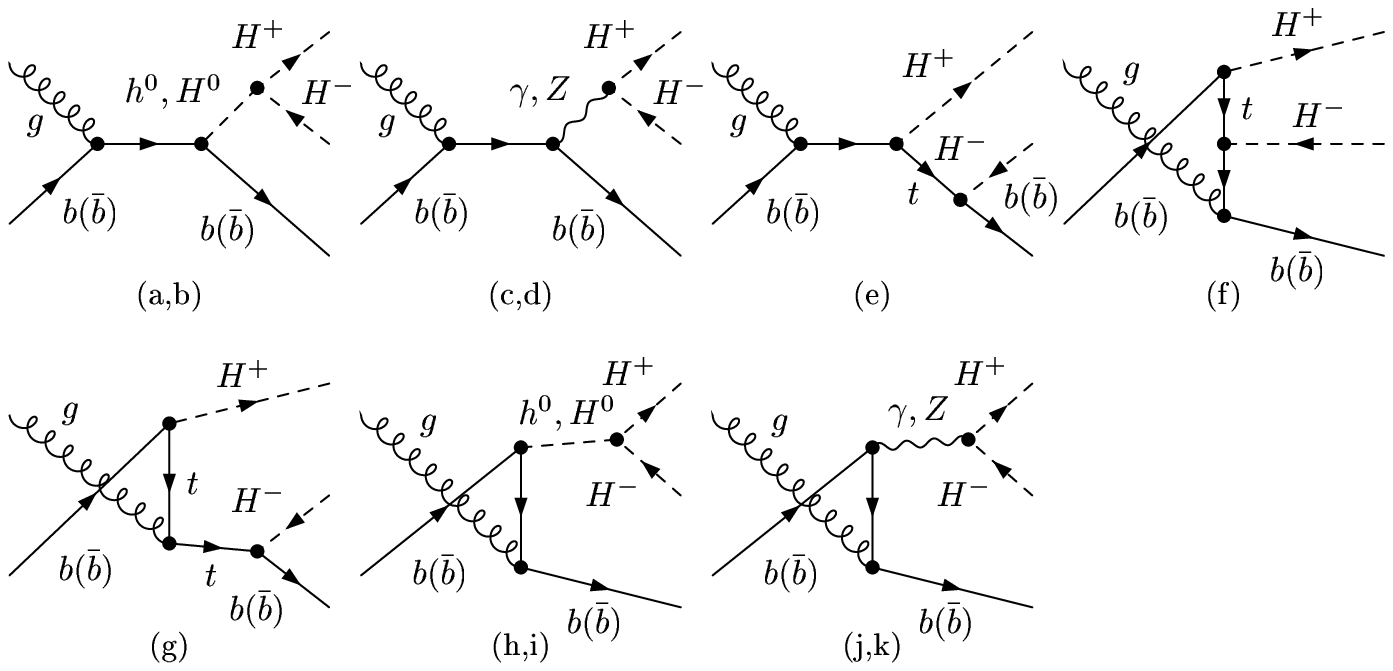}}
\caption{The Feynman diagrams for $g b (\bar{b}) \to H^+ H^- + b
(\bar{b})$.}
\end{figure}

\par

The contributions from these processes only contain the initial
state collinear singularities. Using the method described above,
we split the phase space into two regions: collinear region and
non-collinear region.
\begin{eqnarray}
\hat{\sigma}^R_{b}(g b(\bar{b}) \to H^+H^-+b(\bar{b})) =
\hat{\sigma}^C_{b}(g b(\bar{b}) \to H^+H^-+b(\bar{b}))
+\hat{\sigma}^{\overline{C}}_{b}(g b(\bar{b}) \to
H^+H^-+b(\bar{b})) \nb \\
\end{eqnarray}
Also $\hat{\sigma}^{\overline{C}}_{b}$ is finite and can be
evaluated in four dimensions using standard Monte Carlo
techniques. The differential cross section $d\sigma^C_{b}$ can be
written as
\begin{eqnarray}
\label{collinear-b} d\sigma^{C}_b&=&d\hat{\sigma}^0
\left[\frac{\alpha_s}{2 \pi} \frac{\Gamma(1-\epsilon)}{\Gamma(1-2
\epsilon)}\left(\frac{4 \pi
\mu_r^2}{\hat{s}}\right)^{\epsilon}\right]
(-\frac{1}{\epsilon})\delta_c^{-\epsilon} [2
P_{bg}(z,\epsilon)G_{g/P}(x_1/z)G_{b/P}(x_2) \nb \\
&+& (x_1\leftrightarrow
x_2)]\frac{dz}{z}(\frac{1-z}{z})^{-\epsilon}dx_1dx_2.
\end{eqnarray}
with
\begin{eqnarray}
P_{bg}(z,\epsilon)&=&P_{bg}(z)+ \epsilon P'_{bg}(z), \nb \\
P_{bg}(z)&=&\frac{1}{2}[z^2+(1-z)^2],~~~~~~ P'_{bg}(z) = -z(1-z).
\end{eqnarray}

\subsection{NLO QCD Corrected Cross Section for $pp \to b\bar{b} \to
H^+H^-+X$}

After adding the renormalized virtual corrections and the real
corrections, the partonic cross sections still contain the
collinear divergences which can be absorbed into the redefinition
of the distribution functions at NLO. Using the $\overline{\rm
MS}$ scheme, the scale dependent NLO parton distribution functions
are given as \cite{Harris}
\begin{eqnarray}
G_{i/P}(x,\mu_f)=G_{i/P}(x)+\sum_j(-\frac{1}{\epsilon})\left[\frac{\alpha_s}{2
\pi} \frac{\Gamma(1-\epsilon)}{\Gamma(1-2 \epsilon)}\left(\frac{4
\pi
\mu_r^2}{\mu_f^2}\right)^{\epsilon}\right]\int^1_z\frac{dz}{z}P_{ij}(z)G_{j/P}(x/z).
\nb \\
\end{eqnarray}
By using above definition, we get a NLO QCD parton distribution
function counter-terms which are combined with the collinear
contributions (Eq.(\ref{collinear-g}) and Eq.(\ref{collinear-b}))
to result in the $O(\alpha_s)$ expression for the remaining
collinear contributions:
\begin{eqnarray}
\label{collinear cross section} d\sigma^{coll}&=&d\hat{\sigma}^0
\left[\frac{\alpha_s}{2 \pi} \frac{\Gamma(1-\epsilon)}{\Gamma(1-2
\epsilon)}\left(\frac{4 \pi
\mu_r^2}{\hat{s}}\right)^{\epsilon}\right]\{ 2
\tilde{G}_{b/P}(x_1,\mu_f)G_{\bar{b}/P}(x_2,\mu_f)
\nb \\
&+& 2[\frac{A_1^{sc}(b \to b g)}{\epsilon}+A_0^{sc}(b \to b
g)]G_{b/P}(x_1,\mu_f)G_{\bar{b}/P}(x_2,\mu_f) \nb \\
&+& (x_1 \leftrightarrow x_2)\}dx_1dx_2,
\end{eqnarray}
where
\begin{eqnarray}
A_1^{sc}(b \to bg)&=& C_F(2 \ln \delta_s+3/2), ~~~A_0^{sc} =
A_1^{sc} \ln(\frac{\hat{s}}{\mu_f^2}),
\end{eqnarray}
and
\begin{eqnarray}
\tilde{G}_{b/P}(x,\mu_f)=\sum_{j=b,g}\int^{1-\delta_s
\delta_{bj}}_x \frac{dy}{y}G_{j/P}(x/y,\mu_f)\tilde{P}_{b j}(y),
\end{eqnarray}
with
\begin{eqnarray}
\tilde{P}_{ij}(y)=P_{ij}
\ln(\delta_c\frac{1-y}{y}\frac{\hat{s}}{\mu_f^2})-P'_{ij}(y).
\end{eqnarray}
We can observe that the sum of the soft (Eq.(\ref{soft cross
section})), collinear(Eq.(\ref{collinear cross section})), and
ultraviolet renormalized virtual correction (Eq.(\ref{virtual
cross section})) terms is finite, i.e.,
\begin{eqnarray}
A^S_2&+&A^V_2=0, \nb \\
A^S_1&+&A^V_1+ 2 A_1^{sc}(b\to bg)=0.
\end{eqnarray}
The final result for the total ${\cal O}(\alpha_s)$ correction can
be written as the sum of two terms: a two-body term $\sigma^{(2)}$
and a three-body term $\sigma^{(3)}$.
\begin{eqnarray}
\sigma^{(2)}&=&\frac{\alpha_s}{2 \pi} \int dx_1dx_2d\hat{\sigma}^0
\{ G_{b/P}(x_1,\mu_f)G_{\bar{b}/P}(x_2,\mu_f)[A^S_0+A^V_0+2 A_0^{sc}(b\to bg)] \nb \\
&+& 2 \tilde{G}_{b/P}(x_1,\mu_f)G_{b/P}(x_2,\mu_f)+(x_1
\leftrightarrow x_2 ) \}.
\end{eqnarray}
And
\begin{eqnarray}
\sigma^{(3)}=\int dx_1dx_2
[G_{b/P}(x_1,\mu_f)G_{\bar{b}/P}(x_2,\mu_f)+(x_1 \leftrightarrow
x_2 )]d\hat{\sigma}^{(3)},
\end{eqnarray}
with
\begin{eqnarray}
d\hat{\sigma}^{(3)}=\frac{1}{2\hat{s}_{12}}\{ \int_{\overline{\rm
HC}}\overline{\sum}|M_3(b\bar{b} \to H^+H^-g)|^2 d \Gamma_3+
2\int_{\overline{\rm C}}\overline{\sum}|M_3(gb \to
H^+H^-b)|^2 d \Gamma_3\}.  \nb \\
\end{eqnarray}
Finally, the NLO total cross section for $pp \to b\bar{b} \to
H^+H^-+X$ is given as
\begin{eqnarray}
\sigma^{NLO}=\sigma^{0}+\sigma^{(2)}+\sigma^{(3)}.
\end{eqnarray}

\vskip 5mm
\section{Numerical Results and Discussion}
\par
In this section, we present the numerical results of the cross
section for the charged Higgs boson pair production via bottom
quark fusion at the LHC. In the numerical evaluation, we take the
SM parameters as: $ m_t=178.1$ GeV, $m_Z = 91.1876$ GeV,
$m_{W}=80.425$ GeV and $\alpha_{EW}(m_W) = 1/128$. We use the
two-loop evolution of the strong coupling $\alpha_s(\mu_r)$ with
$\alpha_s(m_Z)=0.1187$ \cite{pdg}. We use the CTEQ6L1 parton
distribution function for the LO cross sections and CTEQ6M for NLO
results \cite{pdfs}. The factorization scale is taken as $\mu_f =
m_{H^{\pm}}/4$ and the renormalization scale is set to be $\mu_r =
m_{H^{\pm}}$ by default unless otherwise stated. We present the
results involving the 'pure' SM-like QCD and the total SUSY QCD
corrections in following two subsections separately.

\subsection{Results Including Only the SM-like QCD Corrections}

In this subsection, we present the cross sections including only
the SM-like QCD corrections. It means we consider this process in
an Two-Higgs-Doublet model without taking SUSY particles into
account. When we calculate the virtual one-loop corrections, we
only include the SM-like QCD one-loop diagrams shown in Fig.2 and
set the renormalization constants in Eq.(\ref{renormalization
constants}) to be their QCD parts. The $\overline{\rm MS}$ bottom
quark mass $\overline{m}_b(\mu_r)$ can be evaluated by using the
one-loop or two-loop renormalization group improved formula with
the bottom quark pole mass taken to be $m_b=4.7$ GeV. They are
expressed as
\begin{eqnarray}
(\overline{m}_b(\mu_r))^{QCD}_{1l} &=& m_b
(1-\frac{4}{3}\frac{\alpha_s(m_b)}{\pi})
\left[\frac{\alpha_s(\mu_r)}{\alpha_s(m_b)}\right]^{(c_0/b_0)}\, , \\
(\overline{m}_b(\mu_r))^{QCD}_{2l} &=& m_b
(1-\frac{4}{3}\frac{\alpha_s(m_b)}{\pi})
\left[\frac{\alpha_s(\mu_r)}{\alpha_s(m_b)}\right]^{(c_0/b_0)}\left[
1+\frac{c_0}{b_0}(c_1-b_1)\left(
\alpha_s(\mu_r)-\alpha_s(m_b)\right) \right] , \nb \\
\end{eqnarray}
where
\begin{eqnarray}
b_0&=&\frac{1}{4 \pi }
(\frac{11}{3}N-\frac{2}{3}n_f),~~~c_0=\frac{1}{\pi}, \\
b_1&=& \frac{1}{2 \pi} \frac{51 N -19 n_f}{11 N - 2 n_f},~~~c_1
=\frac{1}{72 \pi}(101 N - 10 n_f)\, .
\end{eqnarray}
where $N(=3)$ is the number of colors and $n_f(=5)$ is the number
of active light flavors. In our calculation we use
$(\overline{m}_b(\mu_r))^{QCD}_{1l}$ to evaluate the LO cross
sections and $(\overline{m}_b(\mu_r))^{QCD}_{2l}$ for the NLO
cross sections.

\begin{figure}[htbp]
\centering
\scalebox{0.75}{\includegraphics*[20,75][640,400]{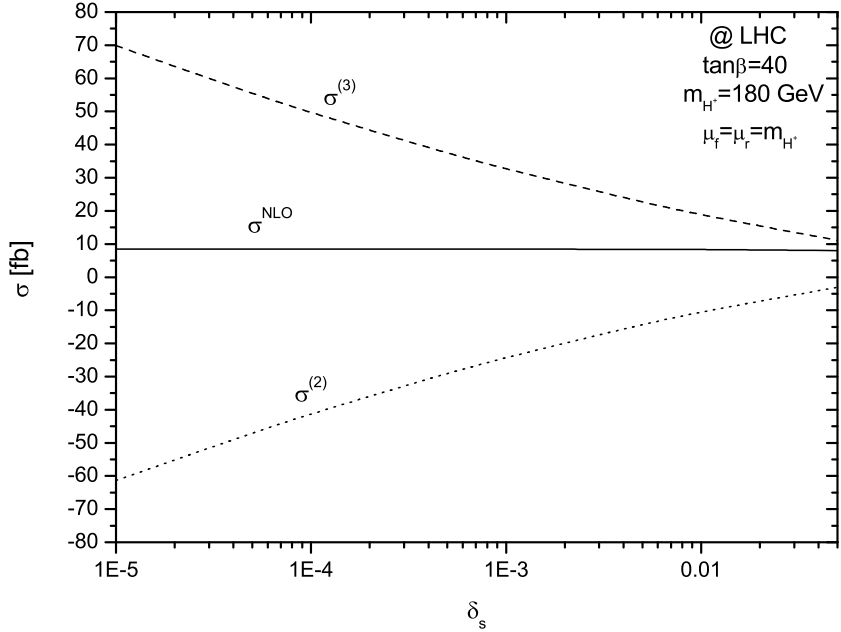}}
\caption{The dependence of the cross sections on the cutoff
$\delta_s$}
\end{figure}

\begin{figure}[htbp]
\centering
\scalebox{0.75}{\includegraphics*[20,75][640,400]{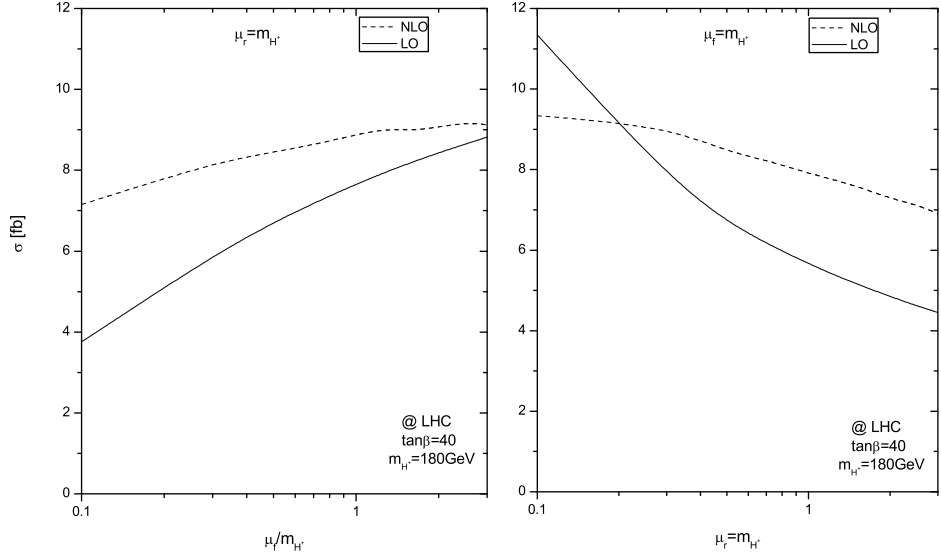}}
\caption{The dependence of the cross sections on factorization
scale $\mu_f$ and renormalization scale $\mu_r$.}
\end{figure}

Fig.6 shows that our NLO QCD result does not depend on the
arbitrary cutoffs $\delta_s$ and $\delta_c$ of the two cutoff
phase space slicing method. The two-body($\sigma^{(2)}$) and
three-body($\sigma^{(3)}$) contributions and the NLO cross section
($\sigma^{NLO}$) are shown as a function of the soft cutoff
$\delta_s$ with the collinear cutoff $\delta_c=\delta_s/50$.
$\tan\beta=40$, $m_{H^{\pm}}=180$ GeV and
$\mu_f=\mu_r=m_{H^{\pm}}$. We can see the NLO cross section
$\sigma^{NLO}$ is independent of the cutoffs. In the following
numerical calculations, we take $\delta_s=10^{-4}$ and
$\delta_c=\delta_s/50$.
\par
In Fig.7, we show the dependence of the total cross section for
$pp \to b\bar{b} \to H^+H^-+X$ on the renormalization scale and
the factorization scale with $\tan\beta=40$, $m_{H^{\pm}}=180$
GeV. In the left plot of Fig.7, the renormalization scale is taken
to be $\mu_r=m_{H^{\pm}}$ while the factorization scale varies in
the region $0.1 m_{H^{\pm}} \sim 3 m_{H^{\pm}}$. In the right plot
of Fig.7, we take $\mu_f=m_{H^{\pm}}$ and $0.1 m_{H^{\pm}} < \mu_r
< 3 m_{H^{\pm}}$. A significant reduction of the scale dependence
for the NLO cross sections can be observed, thus the reliability
of the NLO QCD predictions has been improved substantially.

\begin{figure}[htbp]
\centering
\scalebox{0.75}{\includegraphics*[20,75][640,400]{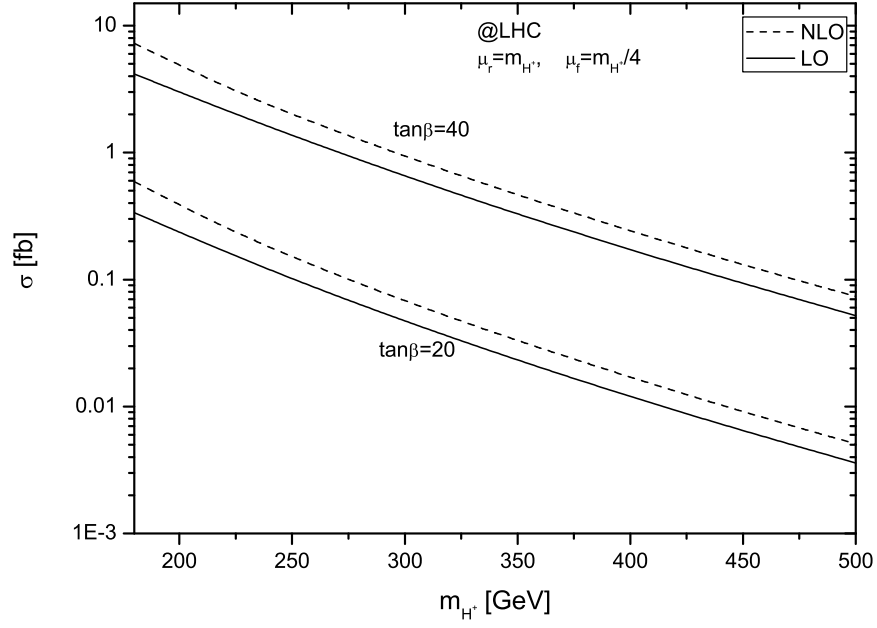}}
\caption{The dependence of the cross sections on $m_{H^{\pm}}$.}
\end{figure}

\begin{figure}[htbp]
\centering
\scalebox{0.75}{\includegraphics*[20,75][640,400]{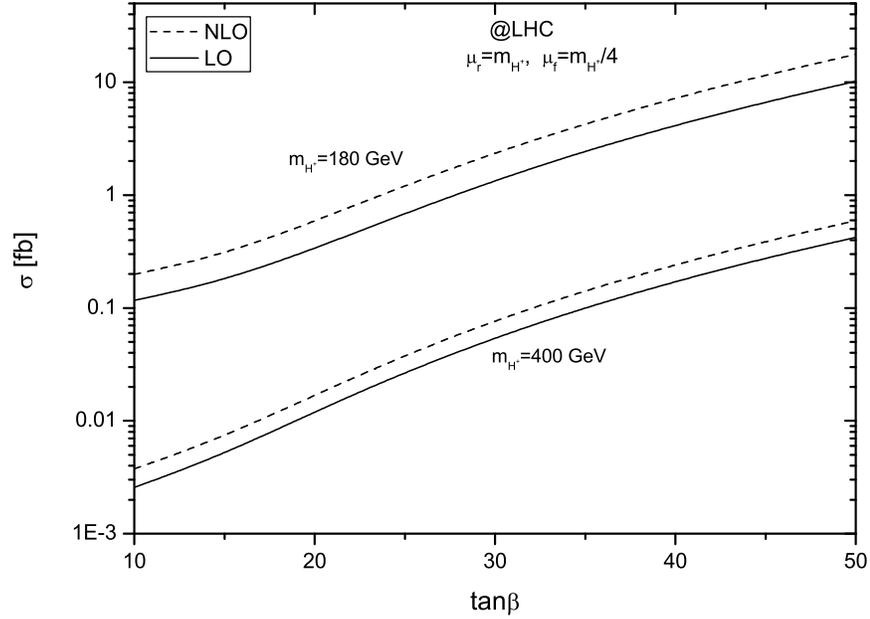}}
\caption{The dependence of the cross sections on $\tan \beta$.}
\end{figure}

\par
Fig.8 shows the dependence of the LO and NLO cross sections for
$pp \to b\bar{b} \to H^+H^-+X$ on the charged Higgs mass
$m_{H^{\pm}}$. The values of $\tan \beta$ are taken to be 40
(upper lines) and 20 (lower lines). $m_{H^{\pm}}$ varies from 180
GeV to 500 GeV. The total cross section can reach 10 $fb$ for
small values of $m_{H^{\pm}}$ with very large $\tan \beta$. In
Fig.9, the dependence of the $pp \to b\bar{b} \to H^+H^-+X$ cross
sections on $\tan \beta$ are studied for $m_{H^{\pm}} = 180$ GeV
and 400 GeV. The cross sections increase rapidly with the
increment of $\tan \beta$ from 10 to 50.

\vskip 5mm
\subsection{Results Including the Total SUSY QCD
Corrections}

In this subsection, we present the cross sections including all
the NLO SUSY QCD corrections. The relevant SUSY parameters in our
calculation are: the parameters
$M_{\tilde{Q},\tilde{U},\tilde{D}}$ and $A_{t,b}$ in squark mass
matrices, the higgsino mass parameter $\mu$ and the mass of the
gluino $m_{\tilde{g}}$. The squark mass matrix is defined as
\begin{eqnarray}
{\cal M}_{\tilde{q}}^2 &=& \left( \begin{array}{cc} m^2_{\tilde{q}_L} & a_q m_q \\
a_q m_q & m^2_{\tilde{q}_R}
\end{array}\right)
\end{eqnarray}
with
\begin{eqnarray}
m^2_{\tilde{q}_L}&=&M_{\tilde{Q}}^2 + m_q^2 + m_Z^2 \cos 2\beta (I^q_3 - e_q \sin^2 \theta_W), \nb \\
m^2_{\tilde{q}_R}&=&M_{\{\tilde{U},\tilde{D}\}}^2 + m_q^2 + m_Z^2
\cos 2\beta e_q \sin^2 \theta_W   \nb \\
a_q&=&A_q-\mu \{\cot\beta,\tan\beta\},
\end{eqnarray}
for \{up, down\} type squarks. $I_3^q$ and $e_q$ are the third
component of the weak isospin and the electric charge of the quark
$q$. The chiral states $\tilde{q}_L$ and $\tilde{q}_R$ are
transformed into the mass eigenstates $\tilde{q}_{1}$ and
$\tilde{q}_{2}$:
\begin{equation}
\left( \begin{array}{cc} \tilde{q}_{1} \\ \tilde{q}_{2}
\end{array}
\right) = R^{\tilde{q}}\left( \begin{array}{cc} \tilde{q}_L \\
\tilde{q}_R \end{array} \right),~~R^{\tilde{q}} = \left(
\begin{array}{cc}\cos\theta_{\tilde{q}} &
 \sin\theta_{\tilde{q}} \\-\sin\theta_{\tilde{q}} & \cos\theta_{\tilde{q}} \end{array}
 \right).
\end{equation}
Then the mass eigenvalues $m_{\tilde{q}_{1}}$ and
$m_{\tilde{q}_{2}}$ are given by
\begin{eqnarray}
\left( \begin{array}{cc} m^2_{\tilde{q}_1} & 0 \\
0 & m^2_{\tilde{q}_2}\end{array}\right)=R^{\tilde{q}}{\cal
M}_{\tilde{q}}^2(R^{\tilde{q}})^{\dagger}
\end{eqnarray}
For simplicity, we assume $M_{\tilde{Q}}= M_{\tilde{U}}=
M_{\tilde{D}}=A_t=A_b=m_{\tilde{g}}\equiv M_{SUSY}=500$ GeV,
$\mu=200$ GeV.
\par
In the MSSM, the counter term of $m_b$ can be very large due to
the 'pure' SUSY QCD (gluino-mediated) diagram for large values of
$\tan \beta$. The gluino-mediated contributions can be absorbed
into the tree-level Yukawa couplings \cite{eberl}. In such a way
we obtain the $\overline{\rm MS}$ bottom quark mass
$\overline{m}_b$ including the total SUSY QCD contributions,
\begin{eqnarray}
\label{mbsqcd} (\overline{m}_b(\mu_r))^{SQCD} =
(\overline{m}_b(\mu_r))^{QCD}-\frac{\alpha_s}{4 \pi} C_F m_b
\{\sum^{2}_{i=1} [B^{fin}_1 -
\frac{m_{\tilde{g}}}{m_b}\sin(2\theta_{\tilde{b}})(-1)^i
B^{fin}_0](m^2_b,m^2_{\tilde{g}},m^2_{\tilde{b}_i}) \} \nb \\
\end{eqnarray}
where the notations $B^{fin}_1$ and $B^{fin}_0$ denote the
operations of taking the finite parts of the two-point integral
functions. The $\mu_r$ dependence of $B$ functions cancels after
summing over the sbottom index $i=1,2$.  As shown in a series
papers, the SUSY QCD contributions to the bottom quark running
mass can be written as \cite{susymb}
\begin{eqnarray}
(\overline{m}_b(\mu_r))^{SQCD} \simeq
\frac{(\overline{m}_b(\mu_r))^{QCD}}{1+\Delta m_b},
\end{eqnarray}
where
\begin{eqnarray}
\label{susy mb} \Delta m_b = \frac{2 \alpha_s}{3 \pi} \mu
m_{\tilde{g}}\tan \beta
I(m_{\tilde{b}_1},m_{\tilde{b}_2},m_{\tilde{g}}),
\end{eqnarray}
with
\begin{eqnarray}
I(a,b,c)=\frac{1}{(a^2-b^2)(b^2-c^2)(a^2-c^2)}(a^2b^2\log
\frac{a^2}{b^2} + b^2c^2\log \frac{b^2}{c^2} + c^2a^2\log
\frac{c^2}{a^2}).
\end{eqnarray}
\par
Indeed, if all supersymmetry breaking mass parameters and $\mu$
are of equal size, one get an interesting limit of Eq.(\ref{susy
mb})\cite{susymb1},
\begin{eqnarray}
\label{approx mb} \Delta m_b = sign( \mu) \frac{\alpha_s
(Q=M_{SUSY})}{3 \pi} \tan \beta.
\end{eqnarray}
We can see $\Delta m_b$ does not decouple in the limit of large
values of the supersymmetry breaking masses. The sign of $\mu$ is
the decisive factor in determining whether the 'pure' SUSY QCD
corrections will enhance or suppress the cross section for the
process of $pp \to b\bar b \to H^+H^-+X$.

\begin{figure}[htbp]
\centering
\scalebox{0.75}{\includegraphics*[20,75][640,400]{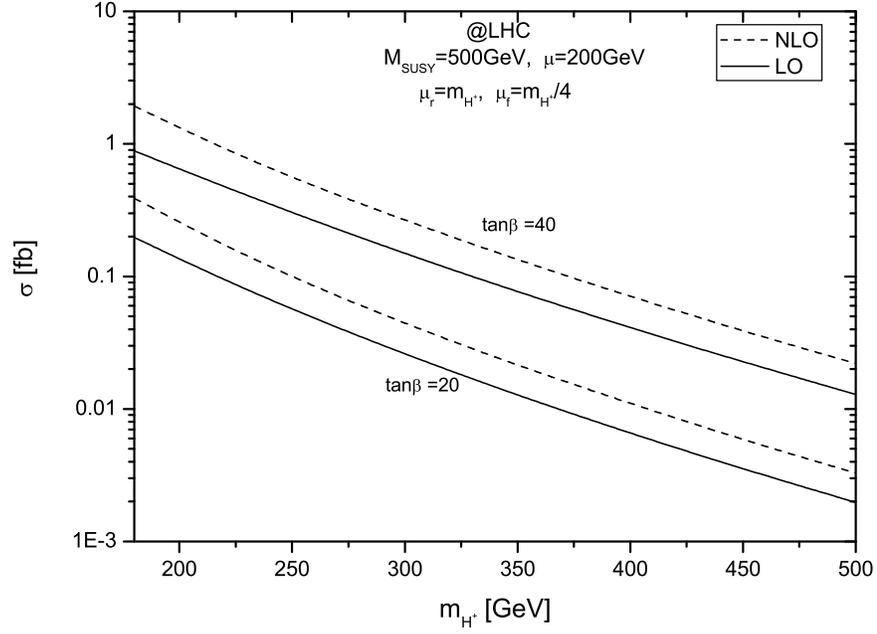}}
\caption{The dependence of the cross sections on $m_{H^{\pm}}$
including the total SUSY QCD corrections.}
\end{figure}

\begin{figure}[htbp]
\centering
\scalebox{0.75}{\includegraphics*[20,75][640,400]{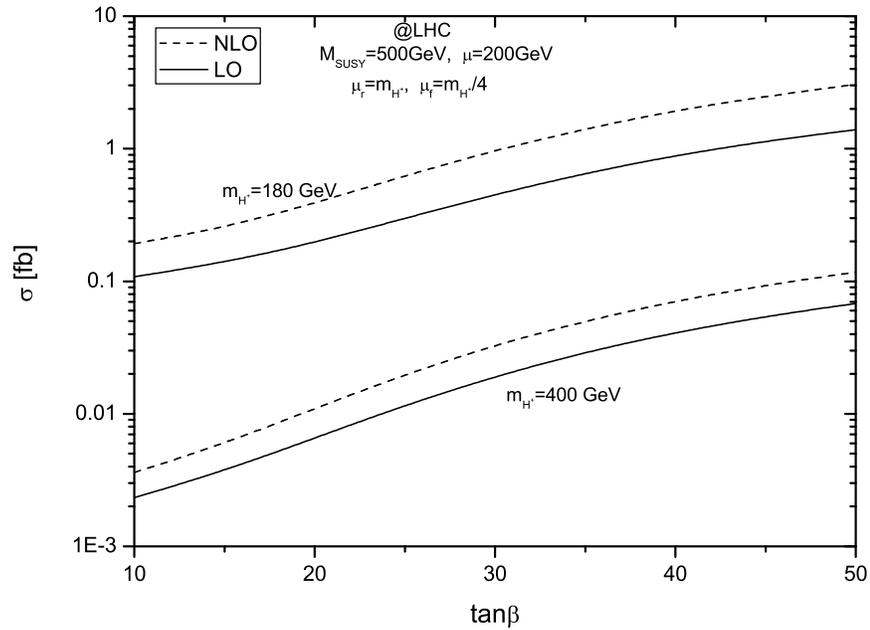}}
\caption{The dependence of the cross sections on $\tan \beta$
including the total SUSY QCD corrections.}
\end{figure}

\par
In our calculations we use Eq.(\ref{mbsqcd}) to calculate the
$\overline{\rm MS}$ running bottom quark mass. The large SUSY QCD
corrections are absorbed in the tree-level bottom Yukawa coupling,
and we use the bottom quark mass counter term defined in
Eq.(\ref{renormalization constants}) to avoid double counting. In
Fig.10, we show the dependence of the LO and NLO cross sections on
the charged Higgs mass $m_{H^{\pm}}$ which is same as Fig.8 but
including the total SUSY QCD corrections. The curves in Fig.10 are
much lower than those in Fig.8, because after including the 'pure'
SUSY QCD contributions $(\overline{m}_b(\mu_r))^{SQCD}$ is much
smaller than $(\overline{m}_b(\mu_r))^{QCD}$, assuming the sign of
$\mu$ is positive and $\tan \beta$ is large (see Eq.(\ref{approx
mb})). In Fig.11, we plot the $\tan \beta$ dependence of the cross
section including the total SUSY QCD contributions.

\par
In summary, we have studied the production of the charged Higgs
boson pair via the bottom quark fusion in the MSSM including the
NLO QCD contributions at the LHC. With very large values of
$\tan\beta$, $b\bar b \to H^+H^-$ subprocess can become the
dominant mechanism in the charged Higgs boson pair production at
the LHC. The numerical results of the cross sections show that the
NLO SUSY QCD corrections are generally significant. We find also
that the $\overline{\rm MS}$ bottom quark mass $\overline{m}_b$
will receive large corrections from the 'pure' SUSY QCD
contributions which will significantly suppress or enhance the
cross section depending on the sign of the higgsino mass parameter
$\mu$.  The NLO QCD corrections can significantly reduce the
dependence of the cross sections on the renormalization and
factorization scales.

\vskip 5mm
\noindent{\large\bf Acknowledgments:} This work was
supported in part by the National Natural Science Foundation of
China and special fund sponsored by China Academy of Science.

\end{document}